\def\year{2018}\relax
\documentclass[letterpaper]{article} 
\usepackage{aaai18}  
\usepackage{times}  
\usepackage{helvet}  
\usepackage{courier}  
\usepackage{url}  
\usepackage{graphicx}  
\usepackage{csquotes}
\begin{document}

\title{Dropout Model Evaluation in MOOCs}

\author{Josh Gardner, Christopher Brooks\\
School of Information\\
The University of Michigan\\
jpgard@umich.edu, brooksch@umich.edu
}

\maketitle
\begin{abstract}

The field of learning analytics needs to adopt a more rigorous approach for predictive model evaluation that matches the complex practice of model-building. In this work, we present a procedure to statistically test hypotheses about model performance which goes beyond the state-of-the-practice in the community to analyze both algorithms and feature extraction methods from raw data. We apply this method to a series of algorithms and feature sets derived from a large sample of Massive Open Online Courses (MOOCs). While a complete comparison of all potential modeling approaches is beyond the scope of this paper, we show that this approach reveals a large gap in dropout prediction performance between forum-, assignment-, and clickstream-based feature extraction methods, where the latter is significantly better than the former two, which are in turn indistinguishable from one another. This work has methodological implications for evaluating predictive or AI-based models of student success, and practical implications for the design and targeting of at-risk student models and interventions.

\end{abstract}

\section{Introduction}

Building predictive models of student success has emerged as a core task in the fields of learning analytics and educational data mining. As MOOCs have grown, so has this need for effective and reliable machine learning models of complex student behavior patterns which identify students likely to drop out in order to provide appropriate interventions or support. The present work is concerned with building and evaluating models to address the following task:

\begin{displayquote}
    \textbf{The Model Selection Task (MST)}: Given the full set of learner data in $N$ courses up to time $t$, find the best of $k$ methods to predict learner dropout at time $t+1$.
\end{displayquote}

More casually, we are interested in determining the best model to use based on Coursera data exports in order to predict learner dropout (no further engagement) at a given week. We are especially interested in building these models for the early weeks of the course: for example, predicting likely dropouts in the third week given data from weeks one and two. We leave the development and evaluation of interventions based on these models to future work.

The process of building such models involves several stages, including extracting structured data from the raw, semi-structured MOOC data (clickstream server logs, database tables, etc.); selecting, training, testing, and tuning different algorithms to predict dropout status; and using statistical inference to identify the ``best'' combination of these techniques. Together, these stages profoundly influence the effectiveness of predictive MOOC dropout models. 

There are three problems with prior learning analytics research on this task: (1) this work isolates the steps of the model-building process, e.g., evaluating different approaches to feature extraction or algorithm selection separately without considering their interplay; (2) there exists no consensus on methods for rigorous and reproducible statistical inference for the MST; and (3) this research is often limited to small subsets of students who display the behaviors within the feature set of interest.

In this paper we apply a nonparametric statistical model evaluation procedure to a large sample of MOOC data. In so doing, we demonstrate the importance of this technique in understanding the relative effectiveness of feature sets and algorithms together on the MST. We find that many complex features which have have shown explanatory power within MOOC student subpopulations show poor performance when applied to dropout prediction on the full population of students. A simple set of clickstream-based features demonstrates much stronger performance, with other features not statistically significantly improving results even in combination with clickstream features, calling into question the value of non-clickstream based approaches.

\section{Related Work}

Previous predictive modeling research in MOOCs has evaluated features derived from clickstreams \cite{Brinton2015-bx,Xing2016-le}, forum post text and behavior patterns \cite{Yang2013-zq,Crossley2015-vy}, assignments \cite{Veeramachaneni2014-ug}, higher-order time series representations \cite{Brooks2015-ej}, emotion \cite{Dillon2016-fa}, and social networks \cite{Rose2014-mk}, among many others. In addition, research has applied many modeling algorithms to dropout prediction, including logistic regression \cite{Kizilcec2015-gu,Veeramachaneni2014-ug}, support vector machines \cite{Kloft2014-kb}, tree-based methods \cite{Li2016-tf}, ensembles \cite{Xing2016-le}, and neural networks \cite{Fei2015-ea}. In most cases these examinations (features, modeling algorithms) are conducted independently of each other.

Despite this robust and growing research base, a recent survey of the literature by the authors ($N=86$) indicated that accepted statistical practices for evaluating these models are often neglected by such research \cite{Gardner2017-vc}. In particular, more than half of surveyed work did not utilize any statistical testing for evaluating model performance, despite conducting model comparisons in contexts where such testing is especially critical, such as when model performance is estimated directly on the training set through cross-validation, or when many comparisons were conducted. An additional 5\% utilized paired $t$-tests, which have a demonstrably inflated Type I error rate \cite{Dietterich1998-vh,Nadeau2003-oq} and low replicability when used for model comparison \cite{Bouckaert2004-oy}, leaving such research susceptible to spurious results -- particularly Type I errors. 

Despite these gaps, effective techniques for the MST do exist \cite{Dietterich1998-vh,Provost2001-wj,Nadeau2003-oq,Yildiz2006-kb,Garcia2008-go}. In particular, \cite{Demsar2006-cx} provides a useful methodology for the MST, using a Friedman test paired with a post-hoc Nemenyi test and an information-dense visualization for displaying the results of this procedure: the Critical Difference (CD) diagram. This approach has been utilized in other fields \cite{Madjarov2012-bo}, but has not been applied to the MST in learning analytics, nor has it been used to evaluate \textit{feature extraction} methods in conjunction with predictive algorithms in any application, to the authors' knowledge.

Further, much prior predictive research in MOOCs utilizes subpopulations of learners for whom the features of interest were available, with 46\% of works somehow filtering the population of students, often eliminating as much as 95\% of the learner population from their analysis\cite{Gardner2017-vc}. For example, \cite{Crossley2015-vy} evaluates the predictiveness of clickstream and forum data only for students who both made a forum post and completed an assignment (426 students out of over 48,000 registrants, 13,314 of whom watched at least one video); \cite{Robinson2016-yr} builds a predictive model using students who started in the first two weeks, completed a pre-course survey, saw utility value of course, were fluent in writing English, intend to complete course, and wrote more than one word on survey, less than 5\% of the population. Because these subpopulations are both \textit{small} and \textit{different} across studies, it is difficult to infer the usefulness of such features for predictive modeling on the full learner population in any course, or to compare results obtained on different subpopulations. We argue that the ideal predictive model is one which yields accurate predictions for the \textit{full} learner population, and such work provides only limited evidence to this end.

Our goal in this work is not to build the ultimate predictive model, but to demonstrate that we can compare classes of features and modeling techniques through statistical means, and to use a case study to demonstrate an inferential procedure for identifying features and models which perform better than others for MOOC dropout prediction. While this may seem like something which should be standard practice, it is not. Our procedure demonstrates that while it is difficult to choose a single best classifier (statistically speaking), it is possible to group classifiers by performance, which yields interesting insights into the value of different feature types. 

\section{Methodology}

In this section, we describe the Friedman and Nemenyi two-stage testing procedure and outline why this procedure can be extended to evaluate \textit{features} in addition to algorithms.

\subsection{Friedman and Nemenyi Two-Stage Procedure}

The MST seeks to identify methods for selecting the best of $k > 2$ dropout prediction models for dropout prediction across $N > 1$ MOOCs of varying domains, size, and structure. We implement a procedure from \cite{Demsar2006-cx} to draw statistical inferences about the respective differences in performance across a set of $k = 8$ models, applying it to feature set-algorithm combinations across a set of $N = 31$ MOOCs described below. The procedure consists of two steps: first, a Friedman test, a non-parametric equivalent of the repeated-measures ANOVA, is used to test the null hypothesis that the performances of all models is equivalent. The Friedman test compares the average \textit{rankings} of the $k$ models across each of the $N$ datasets, calculating the Friedman statistic, which measures the probability of the observed rankings under the null hypothesis, $H_0$, of all models having equivalent performance (and therefore equal expected average rankings)

\begin{equation}
    \chi_{k-1}^2 = \frac{12N}{k(k+1)} \left[ \sum_{j}R_j^2-\frac{k(k+1)^2}{4} \right]
    \label{eqn:friedman}
\end{equation}

where $R_i^j$ is the rank of the $j$th of $k$ algorithms on $N$ datasets. The observed value of the Friedman statistic is compared to a critical value for $N$ and $k$ \cite{Friedman1940-bd,Demsar2006-cx}. If $H_0$ is rejected at the selected significance level (e.g. $\alpha = 0.05$), then we proceed to administer the post-hoc Nemenyi test to examine all pairwise comparisons (if $H_0$ is not rejected, we do not have sufficient evidence to conclude that any differences in performance exist). The Nemenyi test is similar to the Tukey test for ANOVA, and uses a critical difference 

\begin{equation}
    CD = q_\alpha\sqrt{\frac{k(k+1)}{6N}}
    \label{eqn:CD}
\end{equation}

as a threshold to determine whether the performance between any two classifiers is significantly different, where $q_\alpha$ is the Studentized range statistic divided by $\sqrt{2}$.

This particular procedure, which we will call the FNP (for Friedman + Nemenyi Procedure), is advantageous for several reasons. First, the FNP directly accounts for the number of comparisons $k$, as well as for the number of datasets $N$ (the number of \textit{observations} of those comparisons). Other testing procedures, including the variants of $t$-tests developed for model evaluation (i.e. \cite{Nadeau2003-oq}), are only calibrated for individual comparisons. Second, the FNP is nonparametric. Similar parametric procedures, such as ANOVA, require assumptions of normality, equal variance, and commensurability of comparisons across datasets, which are frequently violated and difficult to check.\footnote{See \S 3.2.1 of \cite{Demsar2006-cx} for further discussion.} The FNP, in contrast, only requires that the estimates of model performance and the measured rankings they produce are reliable, datasets are independent, sufficient number of trials were conducted to gather accurate estimates of performance (10-fold CV is generally considered adequate), and preferably, that all algorithms were evaluated using the same random samples \cite{Demsar2006-cx}. Our experiment meets these criteria. The FNP makes no further assumptions about the sampling scheme, distribution of the observed classifier performance, or data. The FNP is also preferable to other techniques which only apply comparisons to a single ``baseline'' classifier, such as the Iman and Davenport test \cite{Demsar2006-cx}, because it accounts for all of the pairwise comparisons conducted in the course of an experiment. This allows us to directly identify the statistically indistinguishable family of "best" models instead of creating a procedure to approximate it, as in \cite{Romero2013-xe}.  Finally, the FNP is preferable to the ``choose the best average performance'' approach adopted by much predictive modeling research in learning analytics for the same reasons that a hypothesis test is preferable to a comparison of many sample means: it differentiates between results which may be spurious and those which are likely to indicate a \textit{true} difference in performance under the null hypothesis of equal performance, based on the available data.

\subsection{Extending the Procedure to Features}

The two-stage Friedman with Nemenyi procedure is an accepted approach for evaluating and comparing different classification \textit{algorithms} and various hyperparameter configurations for each \cite{Demsar2006-cx,Japkowicz2011-lw,Madjarov2012-bo}. However, algorithm selection and hyperparameter tuning are not the only components of the MST. In many fields (learning analytics included) the feature extraction methods used to extract information from ``raw'' data, such as clickstream files and database tables, are considered at least as important as the prediction algorithm \cite{Veeramachaneni2014-ug,Nagrecha2017-dn}. This process is rarely evaluated in conjunction with the rest of the predictive model, and the FNP has not, to our knowledge, been applied to this end; independent statistical evaluation of feature extraction from raw data using \textit{any} method is rare in machine learning research. However, the FNP can easily be extended to evaluate feature extraction. To do so, we vary feature extraction methods across models, with the results of one more feature extraction methods being fed to various algorithms as a $feature set + algorithm$ combination. In applying this method to both of these model characteristics, we can (1) evaluate feature extraction methods as a testable component of the overall modeling process; (2) capture and evaluate the synergy between these two dimensions of predictive models; and (3) make inferences which fully account for the number of comparisons across all of these elements in the MST instead of conducting feature evaluation and model evaluation independently.

\section{Experiment}

For each week of a course we consider the task of predicting, for any student currently active in the course, a binary outcome variable that indicates whether the following week would be their final week of activity or not. We consider this the most useful potential outcome to predict, because it would allow instructors, aides, or even automated platforms to intervene and provide support the following week. We predict on \textit{all} students who have shown any activity in the previous week (students who have shown no activity across any feature set in the previous week are considered dropped out; including those students makes the dropout prediction task too easy, particularly in later weeks of the course). We conduct this evaluation after the first four weeks of each course in an effort to identify at-risk students during the early stages of a course, when most dropouts occur.

\subsection{Dataset}

The data are the raw clickstream files and mySQL data exports from 31 offerings of 5 unique courses offered by the University of Michigan on Coursera. Details about each course are listed in Table \ref{course-table}. Each course utilized each data source (collected clickstream data, utilized discussion forums, and administered quizzes).

\begin{table}[t]
  \centering
  \begin{tabular}{p{3.25cm} c c c }
    \hline 
    \textbf{Course} & \textbf{Runs} & \textbf{Duration} & \textbf{Students}\\ \hline
Inside the Internet & 5 & 12 & 24,562 \\
Instruct. Methods in Health Prof. Education & 5 & 12 & 5,413 \\
Intro Finance & 8 & 16 & 175,532 \\
Intro Thermodynamics & 5 & 14 & 13,508 \\
Model Thinking & 8 & 14& 79,894 \\  \hline
\textbf{Total} & 31 & 12--16 & 298,909

  \end{tabular}
  \caption{Courses used for modeling. Duration in weeks. }~\label{course-table}
\end{table}

\subsection{Features}

From course clickstream files and database tables, we extract and evaluate three of the dominant feature sets which collectively represent the data sources used in most MOOC dropout models \cite{Gardner2017-vc}, detailed in Table \ref{feature-table}:

\begin{itemize}
    \item \textbf{Clickstream Features:} Extracted directly from the course clickstream file, these represent data about students' browsing behavior and utilization of course resources, including viewing videos, forum posts, and other course pages. The simple counting-based features we used are common and have been shown to be effective predictors \cite{Kloft2014-kb,Xing2016-le}.
    \item \textbf{Assignment/Academic Features:} Extracted from mySQL database exports, these capture information about student submission and performance on course assignments. This feature set was inspired by the most predictive features identified in \cite{Veeramachaneni2014-ug}, but due to differences in the underlying data available, some features were modified.
    \item \textbf{Forum/NLP Features:} Extracted from mySQL database exports, these represent the language and activity patterns from course discussion forums. We utilize a series of features that previous research identified as robust and replicable \cite{Andres2016-ju,Crossley2015-vy,Yang2013-zq}. Features relevant to longer writing assignments and not short discussion forum posts were excluded, and others (i.e., sentiment features) added to reflect unique characteristics of forum posts in MOOCs as compared to traditional writing assignments \cite{Wen2014-pv}.
\end{itemize}

We selected these features for several reasons. (1) Each feature set required extracting information from a different raw data source. Researchers and institutions are often faced with the difficult task of identifying the feature extraction methods most useful in their predictive models, and this setup was intended to inform such decisions. Furthermore, because these data sources are common to all major MOOC platforms, they are generalizable and relatively straightforward to extract regardless of course platform, content, or design. (2) The features replicate, as far as possible and with some additions, features shown to be effective predictors of MOOC dropout. Despite our effort at replication, we note that the implementations behind these methods \cite{Xing2016-le,Veeramachaneni2014-ug,Andres2016-ju} are all not openly available, which means our work cannot be considered an \textit{exact} replication. The lack of open infrastructure to support feature and model replication is a significant problem for the learning analytics and predictive modeling communities. (3) Assignment and forum features are available only for subsets of the learner population -- i.e., only those who post in forums or complete assignments -- leaving their usefulness in predictive models of the \textit{entire} learner population undetermined, despite an emphasis on these features in prior research discussed above. This comparison allows us to evaluate such features for \textit{all} active learners, independently and in conjunction with other feature sets, and better represents the goals and practical usage of predictive dropout models. 

The feature sets were assembled in a week-level ``appended'' format, where one set of identically-defined features was appended for each week of the course to build a successively ``wider'' dataset.\footnote{As an example, for the forum posts feature, we generated one column of forum posts \textit{per week} prior to the target week: when predicting dropout in week 4, the dataset included columns for forum posts in week 1, week 2, and week 3.} Prior research suggests that this is an effective approach to representing time-series data in a flat format, and that it improves predictive model performance by expanding the feature space and allowing algorithms to capture interactions between variables across time windows \cite{Kloft2014-kb,Xing2016-le}.

\begin{table*}[t!] 
\centering
    \begin{tabular}{p{4.5cm} p{12.5cm}}
     \hline
    \multicolumn{2}{c}{\textbf{Clickstream}} \\
    \hline
    Forum Views & Number of pageviews of forum pages. \\
    Active Days & Number of days for which user registered any clickstream activity (maximum of 7). \\
    Quiz Views & Number of pageviews of quiz attempt pages, as measured by clickstream features. \\
     Exam Views & Number of pageviews of exam-type quiz pages, as measured by clickstream features. \\
     Human-Graded Quiz Pageview & Number of pageviews of human-graded quiz pages, as measured by clickstream features. \\ 
     \hline
    \multicolumn{2}{c}{\textbf{Assignments}} \\
    \hline
    Average Submission Lead Time & Average time between a quiz submission and deadline for all submissions. \\
     Total Raw Points & Sum of total raw points earned on quizzes. \\
     Average Raw Score & Average raw score on all assignments. \\
     Submissions Per Correct & Total submissions divided by total raw points. \\
     Total quiz submissions & Total count of quiz submissions. \\
     Percent of allowed submissions & Total count of quiz submissions as a percent of the maximum allowed submissions. \\
     Percent of max submissions & A student total number of quiz submissions as a percent of the maximum number of submissions made by any student in the course. \\
     Correct submissions percent* & Percentage of the total submissions that were correct. \\
     Change in weekly average* & Difference between current week average and previous week average quiz grade. \\ 
      \hline
    \multicolumn{2}{c}{\textbf{Forum}} \\
    \hline
     Average Post Sentiment & Average net sentiment of posts (positive - negative). \\
     Positive Posts & Number of posts with net sentiment $ \geq 1$ standard deviation above thread average. \\
     Negative Posts & Number of posts with net sentiment $ \leq -1$ standard deviation below thread average. \\
     Neutral Posts & Number of posts with net sentiment within 1 standard deviation of thread average. \\
     Sentiment Relative to Thread & Average of (post sentiment - avg sentiment for thread). \\
     Post Count & Total number of posts and comments. \\
     Threads Started & Total number of threads initiated by student. \\
     Response Count & Count of posts that were responses (immediately following) another users' post. \\
     Unique Words/Bigrams & Count of unique words/bigrams used across all posts. \\
     Flesch Reading Ease & Flesch Reading Ease score, averaged across all posts for that user. \\
     Flesch-Kincaid Grade Level & Flesch-Kincaid grade level, averaged across all posts for that user. \\
     Upvotes Received & Total net upvotes users' posts received (positive - negative) \\
     Direct-Reply Nodes & Count of unique direct-reply connections (student posts are adjacent on thread). \\
     Thread-Reply Nodes & Count of unique thread-reply connections (students posts are on same thread).  \\ \hline 
    \end{tabular}
    \caption{Feature name and definition by category. Each feature is calculated at the student-week level, resulting in $p \cdot n$ features at week $n$ with one observation for each unique student. Sentiment was extracted using the VADER sentiment analyzer \cite{Hutto2014-po}. Features marked with a (*) were calculated by quiz type (homework, quiz, and video), resulting in three different features, one per quiz type, using that definition.} 
    \label{feature-table}
\end{table*}

\subsection{Modeling Algorithms}

This experiment utilized two algorithms: standard classification trees (an implementation of the classical trees from \cite{Breiman1984-or}; hereafter CART) and adaptive boosted trees (\cite{Friedman2000-fu}, hereafter adaboost, which builds iteratively boosted ensembles of small trees, themselves built using the same classical CART method). We selected these classifiers because (1) they can capture complex interactions in high-dimensional data; (2) they make few assumptions about the underlying data distribution; (3) they are scaleable yet perform well when the number of observations is moderate; (4) their performance on real-world classification tasks is generally strong and well-studied; and (5) they produce models or metrics amenable to post-hoc inspection. Additionally, we were faced with feature sets for which $\geq 95\%$ of observations were missing data (for forum and assignment features, for the reasons mentioned above) and required a method for handling these cases. Instead of introducing an imputation method, we selected these two algorithms, which have a built-in and identical method for handling missing data -- using \textit{surrogate variables}  \cite{Therneau2015-am}. This method has the effect of capturing structure in missing data, instead of simply ignoring or replacing missing values, and we believe it is particularly appropriate in this context, where we expect that there is structure and information embedded in this missingness: a certain ``type'' of student may be more inclined not to post in discussion forums; a surrogate variable approach is capable of learning this relationship when other features are available.

Conducting several rounds of hyperparameter tuning stood to substantially increase the number of comparisons conducted in our analysis, and we were only ultimately interested in comparing the ``best'' model of each algorithm and feature type to each other; additional comparisons would only reduce our ability to detect true, statistically significant differences between models. As a result, prior to conducting the comparison below, we used a random 20\% of the course runs to tune hyperparameters for each model-feature set combination. These hyperparameter settings were used in the rest of the comparison.\footnote{Note that subsequent analysis demonstrated that hyperparameter tuning produced minimal changes in model performance and rankings and did not change the overall findings presented below.}

\section{Results}

\begin{figure}[t!] 
    
    \centering
    \textbf{Week 1}
    \includegraphics[height = 2.8cm, width = \columnwidth]{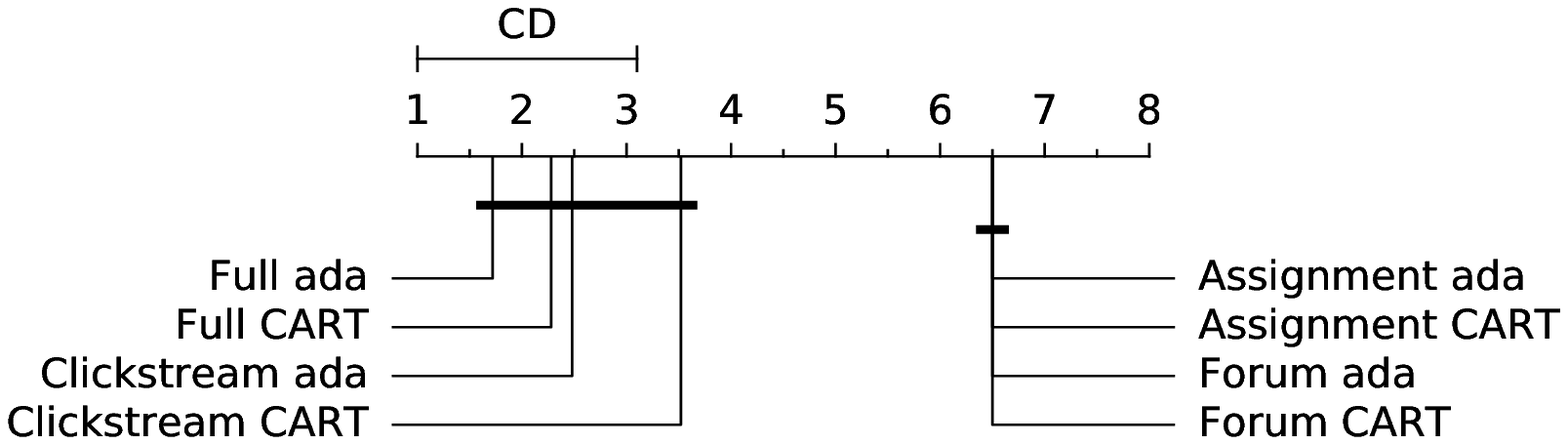}
    \textbf{Week 2}
    \includegraphics[height = 2.8cm, width = \columnwidth]{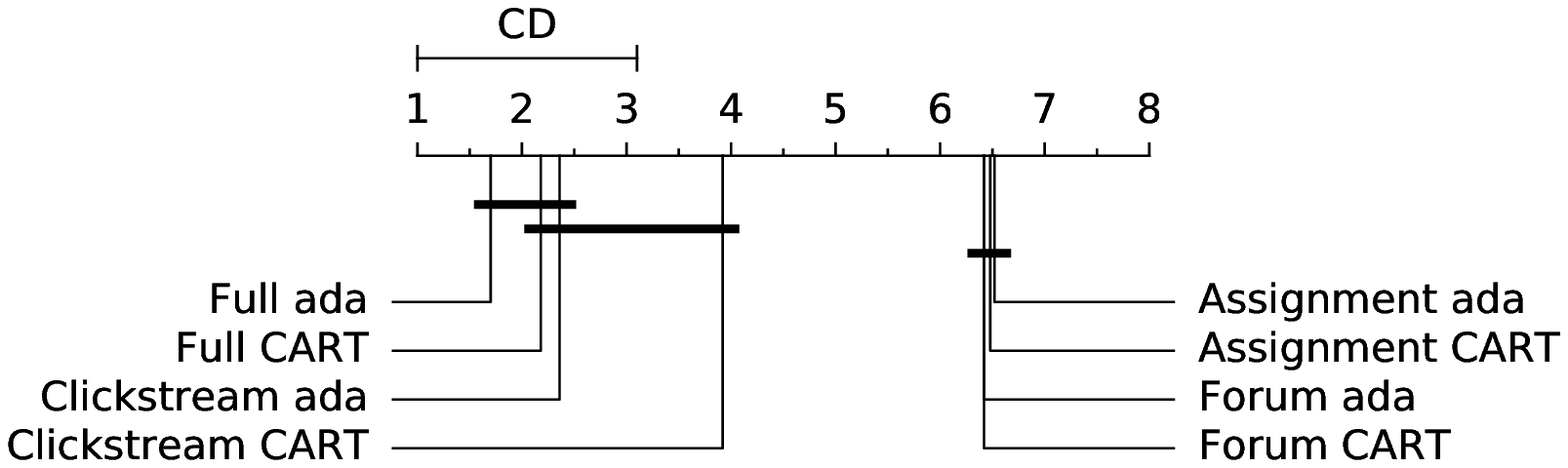}
    \textbf{Week 3}
    \includegraphics[height = 2.8cm, width = \columnwidth]{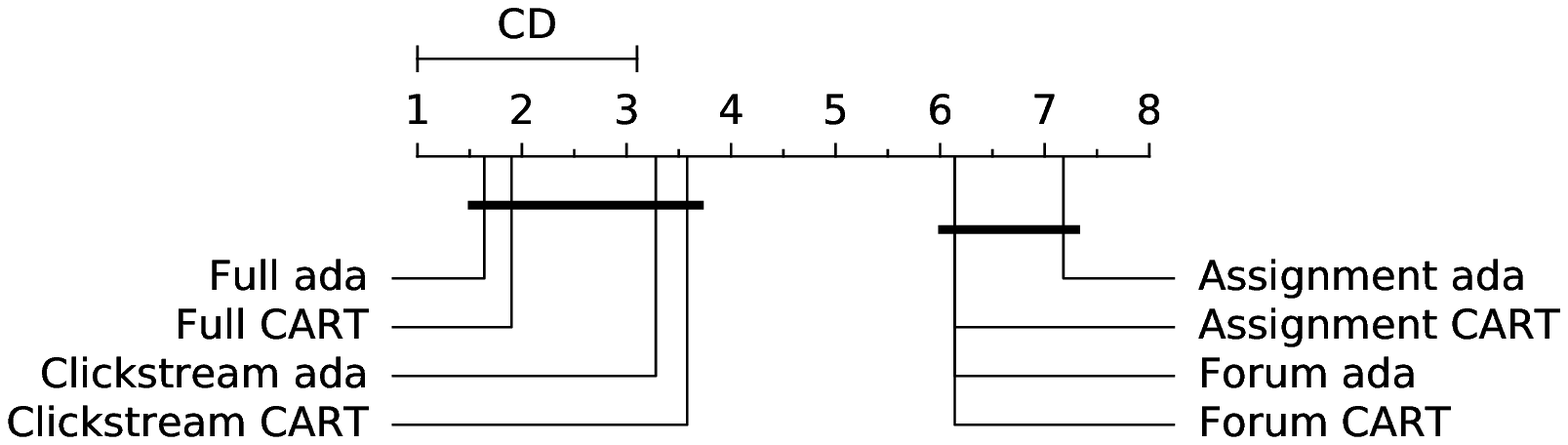}
    \textbf{Week 4}
    \includegraphics[height = 2.8cm, width = \columnwidth]{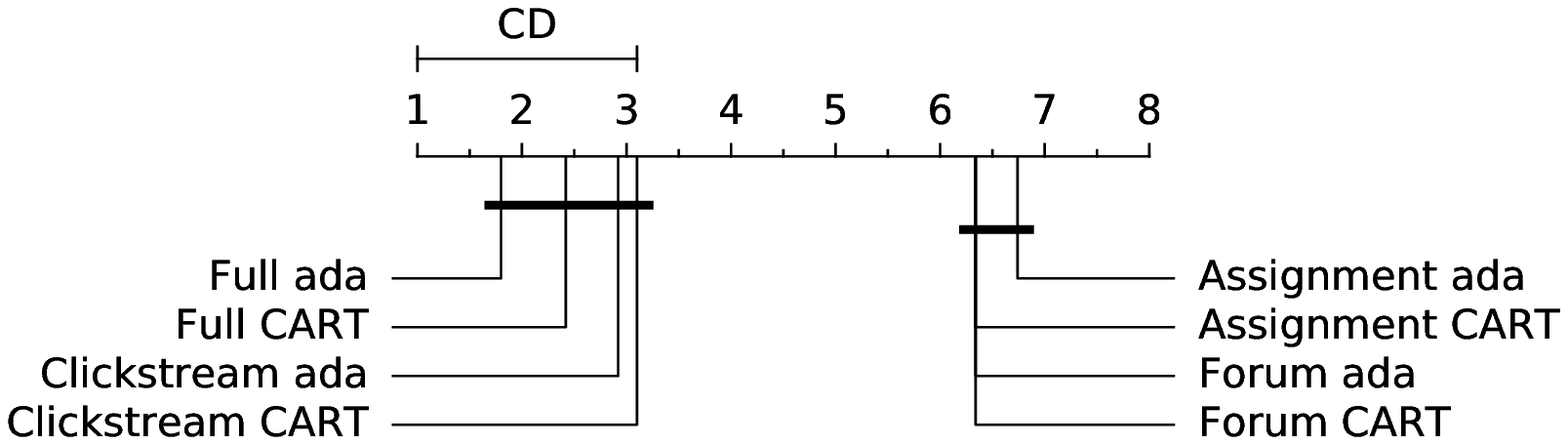}
    \caption{Critical Difference diagrams of results. Each model is plotted according to its average rank; bold CD lines link models statistically indistinguishable at $\alpha = 0.05$. This analysis shows a large, consistent gap between clickstream features and those from assignments and forums.}
    \label{fig:cd-diagrams}
\end{figure}

We evaluate $k = 8$ models for each week, conducting separate experiments to predict dropout after the first, second, third, and fourth week of each course. In this section, we present an our results as an example of how the two-stage Friedman and Nemenyi procedure can be used to address the model selection task (MST). Figure \ref{fig:cd-diagrams} presents the primary results of our analysis in the Critical Difference (CD) diagram discussed below.

\subsubsection{Critical Difference (CD) Diagrams}

The Critical Difference (CD) diagrams represent the results of the Friedman and Nemenyi testing described above. The average rankings of each model are plotted on a number line; models separated by a distance of less than the CD in Equation \ref{eqn:CD} are statistically indistinguishable -- the data is not sufficient to conclude whether they have the same performance -- and are connected by a bold line segment, in effect `linking' them together. Models separated by a distance greater than the CD have a statistically significant difference in performance. We provide an interpretation of the CD diagrams for the results of this experiment below.

\subsection{Model Performance Evaluation}

In all weeks evaluated, the Friedman test rejects $H_0$ of equivalent performance across all models, and so we proceed with the Nemenyi test. Models using quiz and forum features perform significantly worse than other models, and are consistently the poorest-ranking overall predictors of student dropout regardless of the algorithm used. The lack of statistical significance is shown in Figure \ref{fig:cd-diagrams} by the black bar linking all models using quiz and forum features.

The left-hand side of the CD diagram shows that the clickstream features, despite being far less complex, produced far better predictive accuracy and were consistently the highest-performing feature group. The difference between clickstream features and forum/quiz features was statistically significant and remarkably consistent across all four weeks; this is shown in Figure \ref{fig:cd-diagrams} by the lack of a CD line linking algorithms trained with these features. When the engagement features are combined with forum and quiz features to form a ``full'' model, this model achieves better accuracy than the engagement features alone, particularly when used with the more flexible adaboost classifier, but this improvement is never statistically significant. This suggests that the forum and quiz features may contain information relevant to dropout, which interacts with the clickstream data in complex ways that require a highly flexible algorithm to capture.

Changing the classification algorithm had little effect on the accuracy of models utilizing only quiz or forum data, with the various combinations of these features being statistically indistinguishable from each other. This suggests that these sparse features may contain minimal information for those algorithms to capture. Clickstream features, however, did show sensitivity to algorithm selection (though only statistically significant in week 2). We do not present the full ranking results here due to space constraints, and because the CD diagrams summarize this information. Additionally, while the nonparametric testing procedure is specifically intended to avoid the pitfalls inherent in only comparing average accuracy, we note that the absolute difference in average accuracy between the lowest-performing and highest-performing algorithms was between 2\%-6\% each week tested (with accuracies ranging from 69\% to 92.5\%. 

This conclusion -- that the highest-performing model is statistically indistinguishable from several other models in this analysis -- stands in contrast to the practice of much of the prior research surveyed, which simply concludes that the best average performance is the ``best'' model with no statistical testing. This conclusion also highlights the important ways in which the FNP accounts for multiple comparisons, shown in Equation \ref{eqn:CD}, where the critical difference increases as the number of comparisons increase. We note that increasing the number of available datasets would shrink the CD, increasing our ability to detect statistically significant differences between models -- another reason for a shared community infrastructure with respect to MOOC data. Our results do \textit{not} suggest that there is a single ``best'' model among those evaluated, but instead a family of statistically indistinguishable models which may all be effective dropout predctors. From these, we may decide to choose based on other contextual factors: model training and feature extraction time, model interpretability, developer time for writing feature extraction scripts, etc. 

\section{Conclusions}

Our results here demonstrate several important findings that have implications for both the methods and the contents of future research on predictive models of student success. 

First, our results demonstrate that statistical testing of predictive model performance results, rather than simply reporting average performance, can substantially change the inferences we draw from these results. Second, our results show the need for learning analytics researchers to tackle the task of building predictive models for the full population of learners in a course. If we are to build effective predictive models for \textit{all} MOOC learners, future research must accept the challenge of finding features and predictive models effective for this full population. The current experiment demonstrates how model evaluation on this complete population can lead to very different inferences about effective feature extraction and modeling techniques. Third, we demonstrate that this approach can be applied to the complete model-building pipeline, including feature extraction and algorithm selection, in a unified procedure which fully accounts for all experimental comparisons -- a methodological contribution we hope extends beyond learning analytics.

Finally, the specific results of our analysis show that relatively simple clickstream-based features perform much better than complex forum and assignment features only available for small subsets of learners. Additionally, the findings suggest that while forum and assignment features are not by themselves effective predictors, they may add further information to improve upon clickstream-only models, but while this effect was consistently observed, it was not statistically significant with the available data.

\section{Future Research}

We hope that the methods presented in this paper are a catalyst for much-needed future exploration and adoption of methods for statistical evaluation of predictive models in MOOCs. The formation of a consensus on methods for evaluating the entire model-building process will be particularly critical as the technical capabilities of most digital learning platforms approach the point where deployment of such models is natively supported. Future work should extend the current approach to evaluated data imputation and balancing methods, and explore Bayesian methods for model evaluation \cite{Benavoli2016-ff}.

\bibliographystyle{aaai}
\bibliography{EAAI-2018}  

\end{document}